\documentclass[pre,onecolumn,tightenlines,12pt]{revtex4}
\usepackage{graphicx}
\usepackage{dcolumn}
\usepackage{amsmath}
\begin{document}

\title{Random matrix ensembles and the extensivity of the $S_q$ entropy}

\author{A. C. Bertuola$^1$ and M. P. Pato$^2$}

\affiliation{$^{1}$Instituto Federal de Educa\c{c}\~{a}o, 
Ci\^{e}ncia e Tecnologia de S\~{a}o Paulo \\
$^{2}$Instituto de F\'{\i}sica, Universidade de S\~{a}o Paulo
Caixa Postal 66318, 05315-970 S\~{a}o Paulo, S.P., Brazil}

\begin{abstract}

We consider the joint density distribution of the elements
of certain random matrix models which are example of 
globally correlated and asymptotically scale-invariant
distributions. It is shown that in their cases, the 
nonadditive entropy $S_q$ is extensive only when the limit 
$q\rightarrow 1$ is taken. On the other hand, when restriction in the
occupation of the phase space is imposed extensiveness is obtained for
values of the entropic parameter different of one.

\end{abstract}
\maketitle

\section{Introduction}

In recent studies of random matrix theory (RMT) there 
has been an interest in a family of ensembles which at 
the same time generalizes the standard Wigner Gaussian ensemble 
and preserves its invariance under unitary transformations\cite{Mehta}.
This kind of ensembles is generated by superposing 
to the Gaussian fluctuations, an external source of randomness. 
Accordingly, this family has been named disordered 
ensembles\cite{Josue}. It first appeared as a result of applying to RMT
the generalized maximum entropy principle\cite{Bertuola,Raul} 
associated to the nonadditive Tsallis $S_q$  entropy\cite{Tsallis}. 
Generalized in Ref. \cite{Klauder,Josue}, this more general family can be
considered as an instance of the so-called superstatistics\cite{Abul}.     

Given a set of probabilities $p_i$, the Tsallis nonadditive entropy
associated to them is defined as 

\begin{equation}
S_q = \frac{1} {q-1}\left( 1- \sum_i p_{i}^q\right) , \label{1a}
\end{equation}
which, in the $q\rightarrow 1$ limit, recovers the standard
Boltzmann-Shannon (BS) entropy 

\begin{equation}
S_1 =-\sum_i p_i \ln p_i. 
\end{equation}
The nonadditivity of (\ref{1a})
follows since, for a system composed of two independent 
subsystems $A$ and $B$ such that the probabilities $p_i$ are  
product $p_i=p_{m}^{A}p_{n}^{B},$ $S_q\ne S_A + S_B.$ 
Therefore, the definition (\ref{1a}) breaks the additivity
property of the standard entropy. Nevertheless, it has been
claimed\cite{Sato} that, under certain conditions, $S_q,$ 
though nonadditive, asymptotically 
becomes extensive with respect to the definition\cite{penrose}

\begin{equation}
0< \mid \lim_{f\rightarrow \infty} \frac{S_q}{f}\mid < \infty , \label{2a}
\end{equation}
where $f$ is the degree of freedom, while $S_1,$ under the same
circunstance, may or may not be nonextensive. It also has been 
claimed\cite{Sato} that this occurs for
systems in which the phase space is occupied in a scale-invariant
way. We show that, though the 
criterion of scale-invariance is satisfied by the joint density
probability of the matrix elements of the disordered ensembles, 
they only become extensive when the $q\rightarrow 1$ limit is taken.  
On the other hand, if restriction in the variation of the variables 
is introduced by
forcing them to occupy only a fraction of the whole phase space
then, situations occur which made the  nonadditive entropy extensive
for values of the entropic parameter different from one.
Our analysis is peformed, by  considering  joint density distribution of 
random variables continuous and discrete.

\section{Unrestricted occupation of phase space}

Our starting point are two random matrices ensembles: the Gaussian
Wigner ensembles for 
continuous case and for the discrete case, the ensemble of  adjacency 
matrices of the Erd\"{o}s-Renyi model of random graphs. Correlations
are then introduced among the matrix elements by constructing misture 
with some appropriate weight of these two type of matrices. In both 
cases, the scale invariant property is verified independent of the 
number of variables.     

\subsection{Disordered Gaussian ensembles}

The joint density probability distribution of the matrix elements of 
the  RMT Wigner Gaussian ensemble is\cite{Mehta}

\begin{equation}
P_{G} (H )=\left(\frac{\beta}{2\pi}\right)^{f/2}
\exp\left(-\frac{\beta}{2} \mbox{tr} H^{2}\right)  \label{12},
\end{equation}
where $f=N+\beta N(N-1)/2$ is the number of independent matrix elements
for the classes of real matrices ($\beta=1$), complex matrices 
($\beta=2$) and quaternion matrices ($\beta=4$). These three classes
form, respectively, invariant Gaussian ensembles under orthogonal
(GOE), unitary (GUE) and symplectic (GSE)  transformations. The above 
distribution is normalized with respect to the measure 
$dH=\prod_{1}^{N}dH_{ii}\prod_{j>i}\prod_{k=1}^{\beta}\sqrt{2}dH^{k}_{ij}.$  
As matrix elements in  (\ref{12}) are statistically independent
random variables, one should expect that their distribution should 
be additive with respect to the BS entropy and nonadditive 
and nonextensive with respect to $S_q$. In fact, with $q\ne 1$
we have  

\begin{equation}
S_q(f) = \frac{1} {q-1}\left[1-\int dh_1dh_2... dh_f 
\left(\frac{\beta}{2\pi}\right)^{qf/2}
\exp\left(-\frac{q\beta}{2} \sum_{i=1}^f h_i ^{2}\right) \right] ,
\end{equation}
where, with $i=1,2,...,N$, $h_i =H_{ii},$ for the diagonal elements
and, with $i=N+1,N=2,...,f$, $h_i =\sqrt{2}H_{ij}$ for the
off-diagonal forming a set $h$ of $f$ independent random variables. 
Performing the integrals
  
\begin{equation}
S_q(f) = \frac{1} {q-1}\left[1-\left(\frac{q\beta}{2\pi}\right)^{(q-1)f/2}
q^{-f/2}\right] 
\end{equation}
such that it is immediately seen that the ratio $S_q/f$
diverges or vanishes in the limit 
$f\rightarrow \infty.$ On the other hand, if  the limit 
$q\rightarrow 1$ is first taken we get
  
\begin{equation}
S_1(f) = \frac{1}{2}\left(1-\frac{\beta}{2\pi}\right)=S_1(1)f 
\end{equation}
which, as expected, shows additiveness.

Consider now matrices constructed by the relation

\begin{equation}
H(\xi )= \frac{ H_G } {\sqrt{\xi/\bar \xi }} , \label{1}
\end{equation}
where $\xi$ is a positive random variable with a normalized 
density probability distribution $w (\xi )$ with average $ \bar{\xi}.$ 
From (\ref{12}) and (\ref{1}), the joint density probability
distribution 

\begin{equation}
P (H)=
\int d\xi w ( \xi)
\left(\frac{\beta\xi}{2\pi\bar{\xi}}\right)^{f/2}
\exp\left(-\frac{\beta\xi}{\bar{2\xi}}\mbox{tr} H^2  \right) \label{9}  
\end{equation}
follows which, expressed in terms of the reduced set of  
matrix elements, becomes

\begin{equation}
P_f (h_1 , h_2 ,..., h_f )=
\int d\xi w (\xi )\left(\frac{\beta\xi}{2\pi\bar{\xi}}
\right)^{f/2}\exp\left(-\frac{\beta\xi}
{\bar{2\xi}}\sum_{i=1}^{f} h_i^2  \right) . \label{6}  
\end{equation}
Clearly, $h$ is now a set of $f$ correlated variables with
distribution which has the scale-invariance property since removing by 
integration any one of the variables, say $h_k,$ produces the  
same joint distribution $P_{f-1},$ that is

\begin{equation}
\int dh_k P_f (h_1 , h_2 ,..., h_f )= P_{f-1} (h_1 , h_2 ,..., h_{f-1}). 
\label{6b}  
\end{equation} 

The Tsallis entropy of $P_f (h)$ is  by definition

\begin{equation}
S_q (f)= \frac{1} {q-1}\left[1-\int dh_1dh_2... dh_f 
P_{f}^q (h_1 , h_2 ,..., h_f )\right] ,
\end{equation}
which through  the substitution 
$h_{i}^{\prime}=\sqrt{\frac{\beta}{2\pi\bar{\xi}}}h_i$ becomes

\begin{equation}
S_q(f) = \frac{1} {q-1}\left[1-\left(\frac{\beta}{2\pi\bar{\xi}}
\right)^{(q-1)f/2}\int dh_1^{\prime}dh_2^{\prime}... dh_f^{\prime} 
P_{f}^{\prime q} (h_1^{\prime} , h_2^{\prime}  ,..., 
h_f^{\prime})\right] . \label{53}
\end{equation}
If the integral in (\ref{53})
has a finite limit when  $f\rightarrow \infty,$ the entropy will
diverge or vanish for $q\ne 1.$ In this case, extensivity can only happen
if the limit $q\rightarrow 1,$ is first taken or, if the limit 
$f\rightarrow \infty$ is taken keeping fixed the quantity 
 
\begin{equation}
\lambda = \frac{1}{q-1}-\frac{f}{2} . \label{35}
\end{equation}
In this case, concomitantly we have the limits $q\rightarrow 1,$ 
$(q-1)f\rightarrow 2$ and 

\begin{equation}
\lim_{f\rightarrow\infty}\frac{S_q}{f}
=\frac{1}{2}
\left(1-\frac{\beta I}{2\pi\bar{\xi}}
\right).
\end{equation}
are  implied, where $I$ denotes the limiting value of the integral. 

To illustrate the above discussion, take for the weight function 
$w(\xi)$ the gamma distribution 

\begin{equation}
w(\xi)=
\exp(-\xi) \xi^{\bar{\xi} -1} /\Gamma(\bar{\xi})  \label{18}
\end{equation} 
with variance $\sigma_{\xi}=\sqrt{\bar{\xi}}$ of previous studies 
of disorder ensembles. This choice entails power-law behavior for
statistic measures of the ensemble. Substituting (\ref{18})
in the above expressions, the integrals are readily performed such that
(\ref{6}) gives

\begin{equation}
P_f(h;\bar{\xi} )=\left(\frac{\beta}{2\pi\bar{\xi}}
\right)^{\frac{f}{2}}\frac
{\Gamma \left( \bar{\xi}+f/2\right)}{\Gamma \left( \bar{\xi} \right) }
\left(1+\frac{\beta}{2\bar{\xi}}
\sum_{i=1}^f h_i ^{2}\right) ^{-\bar{\xi}-f/2}  \label{22}
\end{equation}
for the ensemble distribution and

\begin{equation}
p(h;\bar{\xi})=\left(\frac{\beta}{2\pi\bar{\xi}}
\right)^{\frac{1}{2}}\frac{\Gamma \left(\bar{\xi} +1/2 \right) }
{\Gamma \left( \bar{\xi}\right)}
\left(1+\frac{\beta}{2\bar{\xi}}
h^{2}\right) ^{-\bar{\xi}-1/2}  \label{280}
\end{equation}
for the density  distribution of a given matrix element.  
Since for large $\left|h\right|,$ 
$p_{\beta}(h;\bar{\xi}) \sim 1/\left|h\right|^{2\bar{\xi}+1},$ 
the distribution (\ref{280}) does not have moments of order superior to
$2\bar{\xi}.$ This fact makes the value $\bar{\xi} =1$ critical since
below it (\ref{280}) does not have second moment\cite{Bertuola}. 

Substituting (\ref{18}) in (\ref{53}), the integrals are
easily performed and the analytic expression  

\begin{equation}
S_q (f) = \frac{1} {q-1}\left[1-\left(\frac{\Gamma(\bar{\xi}+f/2)}
{\Gamma(\bar{\xi})}\right)^{q}\left(\frac{\beta}{2\pi\bar{\xi}}
\right)^{(q-1)f/2}
\frac{\Gamma\left(q\bar{\xi}+(q-1)f/2\right)}{\Gamma\left(
q(\bar{\xi}+f/2)\right)}\right] .  \label{35s}
\end{equation}
for the entropy is obtained.
In order to $S_q $ be extensive, the second term
inside the parenthesis has to become proportional to $f$ for very
large $f$. Replacing the gamma functions by its Stirling
approximation, it is simple to show that
this proportionality does not happen if $q\ne 1 .$ Instead, if
first the limit 
$q\rightarrow 1$ is taken with the degree of freedom $f$
fixed, the BS entropy is

\begin{equation}
S_1(f) = -\ln\left[\frac{\Gamma(\bar{\xi}+f/2)}{\Gamma(\bar{\xi})}
\right]-
\frac{f}{2}\ln(\frac{\beta}{2\pi\bar{\xi}})-(\bar{\xi}+f/2)
\left[\Psi(\bar{\xi})-\Psi(\bar{\xi}+f/2)\right] ,
\end{equation}
where $\Psi(x)$ denotes the 
logarithmic derivative of the gamma function. Taking now the limit 
$f\rightarrow \infty,$ we obtain 

\begin{equation}
S_1(f) = \frac{f}{2}\left[1-\ln(\frac{\beta}{2\pi\bar{\xi}})
-\Psi(\bar{\xi})\right]= fS_1(1) ,
\end{equation}
showing the additiveness of the Boltzmann-Shannon entropy.

Let us now take
the limit $f\rightarrow \infty$ imposing the condition (\ref{35})
with $\lambda= \bar{\xi} $  then the  limit expression 

\begin{equation}
\lim_{f\rightarrow\infty}\frac{S_q}{f}
=\frac{1}{2}
\left(1-\frac{\beta}{2\pi e}
\right)                            \label{35t}
\end{equation}
for the entropy is obtained ($e=2.71828...$) proving its extensivity. 
In Fig. 1, the exact and the asymptotic expressions of the $S_q$ 
entropy are plotted as function of $f.$

\subsection{Disordered random graphs}

A graph is an array of points (nodes) connected by edges. It is  
completely defined by its adjacency matrix, $A,$ whose elements, $A_{ij},$ 
have value $1$($0$) if the pair $(ij)$ of nodes is connected (disconnected).
The diagonal elements are taken equal to zero, i.e. $A_{ii}=0.$ 
Adjacency matrices are real symmetric random matrices for graphs in
which the connections between pairs of nodes are randomly set. 
The classical graph of this kind was 
proposed by  Erd\"{o}s and Renyi (ER) in which pair of nodes are 
independently connected with a fixed probability $p$\cite{Renyi}. 
The properties of this classical model are strongly dependent
on the value of the probability $p$ compared with the total number 
of nodes $N$ and it is usual to assume the scaling 
$p \sim N^{-z}$ ($z>0$). Two important statistical properties in the
studies of graphs are the eigenvalue density of the adjacency
matrix and the degree probability that gives the probability
distribution of the number of connections of a given node. 

For not too small values of $p,$ that is not too rarefied 
ER graphs, the eigenvalue density  
obeys the Wigner semi-circle law\cite{Mehta,Albert} 

\begin{equation}
\rho_{ER}(E,\alpha)=\left\{
\begin{array}{rl}
\frac{1}{2\pi \sigma^2}\sqrt{4N\sigma^2-E^{2}}, &\mbox{if } 
|E|<\sqrt{4N\sigma^2}\\
0, &\mbox{if } |E|>\sqrt{4N\sigma^2}
\end{array}
\right.                     ,
\end{equation}
where

\begin{equation}
\sigma^2=p(1-p) 
\end{equation}
is the variance of the matrix elements. 
Deviations from this law appear\cite{Leticia,Farkas}
if $p\sim 1/N $ ($z\sim 1$).
The degree distribution, that is the number $k_i$ of 
connections of a given node $i$ is obtained from the
adjacency matrix as     
\begin{equation}
k_i=\sum^{N}_{j=1}A_{ij}
\end{equation}
and the probability $P_k$ of having $k_i$ equal to $k$ is

\begin{equation}
P_k =\frac{\left(N-1\right)!}{k!(N-1-k)!}
p^k (1-p)^{N-1-k}. \label{73}
\end{equation}
For large $N,$ it can be assumed that the nodes are statistically
independent in such a way that (\ref{73}) also gives the 
probability of finding the average number of nodes with $k$ 
connections in the graph\cite{Albert}. 

In Ref. \cite{Josue}, it has been shown that the joint 
distribution of the matrix elements of the ER adjacency matrix 
can be written in terms of its trace as 

\begin{equation}
P_{ER}(A,\alpha)=\left[1+\exp(-\alpha)\right]^{-f}
\exp\left(-\frac{\alpha}{2} \mbox{tr} A^{2}\right) \label{17}
\end{equation}
where $f=\frac{N(N-1)}{2}$ is the number of independent 
matrix elements for a graph with $N$ nodes. Eq. (\ref{17})  
shows that the ER model can be considered as a special case
of the RMT Gaussian ensembles in which matrix elements are Bernoulli
random variables. Focussing only on the set 
$a=a_1,a_2,...,a_f$ of $f$ independent elements, 
(\ref{17}) becomes

\begin{equation}
P_{ER}(a)=\left[1+\exp(-\alpha)\right]^{-f}
\exp\left(-\alpha \sum_{i=1}^f a_i \right), \label{17b}
\end{equation}
from which the probability of one given element,
say $a_k,$ is 

\begin{equation}
p(a_k)=\frac{\exp\left(-\alpha a_k\right)}{1+\exp(-\alpha)} 
=\left\{
\begin{array}{rl}
\frac{\exp(-\alpha)}{1+\exp(-\alpha)} ,& \text{if } a_k = 1 \\
\frac{1}{1+\exp(-\alpha)}, & \text{if } a_k = 0 ,
\end{array}   
\right.  . \label{17c}
\end{equation}
(\ref{17c}) implies in the relation 

\begin{equation}
\alpha=\ln(\frac{1}{p}-1). \label{57}
\end{equation}
between the parameter $\alpha$ and the probability $p.$
Since $p$ is defined in the interval $[0,1],$ the correspondent 
domain of variation of $\alpha$ is $(\infty,-\infty).$   

As $a$ is a set of binary discrete variables, the joint 
probability $P_{ER}(a,\alpha)$ has a finite set of $f+1$ possibilities
which can be arranged in the Leibnitz triangle

\begin{center}
\begin{tabular}{c c c c c c c c c c c c }
(f=0)& & & & & &  & & $1$ & & &  \\ 
(f=1)& & & & & &  & $ p $ &  & $1-p$ & &  \\ 
(f=2)& & & & & & $p^2$ &  & $p(1-p)$  &  & $(1-p)^2$ & \\ 
(f=3)& & & & & $p^3$ &  & $p^{2}(1-p)$  &  & $p(1-p)^2$ &  &$(1-p)^3$, \\
\end{tabular} 
\end{center}
which has the property of adding two adjacent entries of
a row, the above entry on the immediate row is obtained. This property
is equivalent to say that the probability distribution,
Eq. (\ref{17b}), is scale-invariant\cite{Sato}.
As the set  $a$ is also statistically independent, 
the BS entropy of its distribution should be extensive. In fact,
the nonadditive entropy of (\ref{17b}) is easily calculated to be 
given by

\begin{equation}
S_q(f) = \frac{1} {q-1}\left[1-\frac
{\left(1+\exp(-q\alpha)\right)^f}
{\left(1+\exp(-\alpha)\right)^{qf}}
\right]   ,
\end{equation}
which divide by $f,$ diverges or vanishes in the limit $f\rightarrow
\infty$ if $q<1$ or   $q>1,$ respectively. On the other hand, 
taking first the limit $q\rightarrow 1,$ we find

\begin{equation}
S_1(f) = f\left[1-\frac{\alpha}
{\left(1+\exp(\alpha)\right)}-\ln\left(1+\exp(-\alpha)\right)
\right]=fS_1(1) ,
\end{equation}
such that the additiviness of BS entropy follows. 

Proceeding as in the Gaussian ensemble, correlations are introduced in
the ER model by defining a disordered model constructed by superposing
with a given weight ER graphs with different probabilities. Explicitly,
we consider a random graph whose adjacency matrix has joint distribution of
elements given by

\begin{equation}
P(A;\alpha)=\int d \xi 
w(\xi)\frac{\exp\left(-\frac{\alpha\xi}{2\bar\xi}\mbox{tr} A^{2}\right)}
{\left[1+\exp(-\alpha\xi/\bar\xi)\right]^{f}}. \label{515}
\end{equation}
Eq. (\ref{515}) extends to discrete variables the formalism 
defined by Eq. (\ref{9}). Again the width of the distribution 
of $w(\xi )$ is a controling parameter 
but here the parameter $\alpha$ defining the mean ER also plays an 
important role. 

As in the case of the Gaussian ensembles, statistics measures of 
the averaged graph, i.e. the generalized model, are obtained as 
averages over the statistics measures of the ER model. 
For instance, the eigenvalue density is given by

\begin{equation}
\rho(E;\alpha)=\int^{\xi_m}_{0}d\xi
w(\xi)\frac{1}{2\pi p(\xi)[1-p(\xi)] }
\sqrt{4Np(\xi)[1-p(\xi)]-E^{2}} \label{518}
\end{equation}
where

\begin{equation}
p(\xi)=\frac{\exp(-\alpha\xi)}{\left[1+\exp(-\alpha\xi)\right] }
\end{equation}
and

\begin{equation}
\xi_m=\frac{2}{\alpha}\cosh^{-1} (\frac{\sqrt{N}}{E});
\end{equation}
while the degree distribution is given by 

\begin{equation}
P_k(\alpha) =\frac{\left(N-1\right)!}{k!(N-1-k)!} 
\int^{\infty}_{0}d \xi w(\xi)
\frac{\exp\left(-\frac{k\alpha\xi}{\bar\xi}\right)}
{\left[1+\exp(-\frac{\alpha\xi}{\bar\xi})\right]^{N-1}}. \label{98}
\end{equation}
In terms of the set $a$ of independent matrix elements, (\ref{515}) 
becomes 

\begin{equation}
P_f (a;\alpha)=\int d \xi 
w(\xi)\left[1+\exp(-\frac{\alpha\xi}{\bar\xi})\right]^{-f}\exp
\left(-\frac{\alpha\xi}{\bar\xi} \sum_{i=1}^f a_i\right). \label{515b}
\end{equation}
Due to the linearity property of integration, the finite number of 
probabilities defined by (\ref{515b}) can also be arranged in a 
Leibnitz triangle in such a way that
the relations between elements of neighboring rows are
preserved. Therefore, this generalization of the ER graph 
introduces correlation among the variables but their joint probability
distribution still satisfies the scale-invariant criterion, namely

\begin{equation}
\sum_{a_k=0}^{1}P_f (a;\alpha)=  P_{f-1} (a;\alpha) . \label{515d}
\end{equation}
The $S_q$ entropy of (\ref{515b}) is 

\begin{equation}
S_q (f) = \frac{1} {q-1}\left[1-
\sum_{{a_1 a_2 ...a_f}} P^q _f (a_1,a_2,...,a_f;\alpha)\right] ,
\end{equation}
or, explicitly, 

\begin{equation}
S_q (f) = \frac{1} {q-1}\left[1-
\sum_{k=0}^f\frac{f!}{k!(f-k)!}\left( \int d \xi 
w(\xi)\left[1+\exp(-\frac{\alpha\xi}{\bar\xi})\right]^{-f}
\exp (-\frac{\alpha\xi k}{\bar\xi})\right)^q\right] . \label{98b}
\end{equation}

To proceed, we choose the
weight distribution $w(\xi)$ to be  given by Eq. (\ref{18}).
In Ref. \cite{Josue} it is shown that with this choice the generalized
model has features of a scale-free graph. In particular, 
the eigenvalue density exhibits a crossover from the Wigner 
semi-circle law to a distribution  highly peaked with heavy
exponential tails.

With this choice, the integral  

\begin{equation}
I_{fk} = \frac{1}{\Gamma(\bar \xi)}
\int_0 ^{\infty} d\xi 
\frac{\exp\left[-(1+\frac{k\alpha}{\bar\xi})\xi\right]\xi^{\bar{\xi} -1}}
{\left[1+\exp(-\frac{\alpha\xi}{\bar\xi})\right]^{f}}, \label{98c}
\end{equation}
in  (\ref{98b}), asymptotically can be calculated in the following
way. We  put the integrand 
in an exponential form $\exp(-F_{fk})$ with

\begin{equation}
F_{fk}(\xi)= (1+\frac{k\alpha}{\bar\xi})\xi-(\bar{\xi} -1)\log\xi
 +f\log\left[1+\exp(-\frac{\alpha\xi}{\bar\xi})\right]
\end{equation}
For  $\bar \xi >1,$ $F_{fk}(\xi)$ has 
a parabolic shape such that a new variable $t$ can be introduced through 
the mapping

\begin{equation}
F_{fk}(\xi)=F_{fk}(\xi_s)+\frac{t^2}{2} ,
\end{equation}
where $\xi_s$ is the minimum value obtained by finding the root of
the equation $F_{fk}^{\prime} (\xi)=0.$ In terms of $t$ the integral becomes 
  
\begin{equation}
I_{fk} =\frac{w(\xi_s)
\exp (-k\alpha\xi_s /\bar\xi)}
{\left[1+\exp(-\alpha\xi_s/\bar\xi)\right]^{f}}
\int^{\infty}_{-\infty}dt(\frac{d\xi}{dt})^{-1}\exp(-\frac{t^2}{2})  .
\end{equation}
Replacing $d\xi/dt$ by its value at $t=0,$ 
the asymptotic approximation

\begin{equation}
I_{fk}  =
\frac{w(\xi_s)
\exp (-k\alpha\xi_s /\bar\xi)}
{\left[1+\exp(-\alpha\xi_s/\bar\xi)\right]^{f}}
\sqrt{\frac{2}{\pi F_{fk}^{\prime\prime}(\xi_s)}} 
\end{equation}
is obtained which replaced in  (\ref{98b}) leads to the asymptotic
approximation 

\begin{equation}
S_q (f)= \frac{1} {q-1}\left[1-
\sum_{k=0}^f\frac{f!}{k!(f-k)!}\left(\frac{w(\xi_s)
\exp (-\alpha\xi_s k/\bar\xi)}
{\left[1+\exp(-\alpha\xi_s/\bar\xi)\right]^{f}}
\sqrt{\frac{2}{\pi G_{fk}^{\prime\prime}(\xi_s)}}
\right)^q\right]   \label{35u}
\end{equation}
for the entropy. In Fig. 2, (\ref{35u}) is plotted as a function of
$f$ imposing the condition (\ref{35}) with ${\bar \xi}=6.$ The linear
increase shows the extensivity of $S_q$ under these condition. 

\section{Restricted occupation of phase space}

The previous section shows that correlations among the
random variables are not sufficient to make the $S_q$ entropy 
of their joint distribution extensive for $q\neq 1$. 
The calculations have been performed allowing the variables 
to freely occupy the phase space in an unrestricted way. In
this section, we again start considering the Wigner and 
the Erd\"{o}s-Renyi models but imposing the condition that the 
random variations of their constituents
are restricted to a region of their phase space. By doing this, we
find situations in which $S_q$, in both cases, becomes extensive for
values of the entropic parameter different from one. The
scale-invariant property is asymptotocally obeyed but in a less
strong form.     

\subsection{Continuous case}

Consider the joint distribution of the matrix elements of the Wigner
ensemble, Eq. (\ref{12}), but with the additional constraint that the
trace is restricted to be less than a maximum value $R^2.$ In terms of
the reduced set $h$ of elements this implies in the 
condition\cite{Bronk}

\begin{equation}
\sum h^2_i<R^2 ,
\end{equation}
such that the elements are forced to be
inside a hypersphere of radius $R.$ As a consequence, the joint 
density distribution has to be renormalized as 

\begin{equation}
P_f (h_1,h_2,...,h_f)=Z_f^{-1}\exp\left(-\frac{\beta}{2}\sum^f _{i=1} 
h^2_i\right) ,
\end{equation}
where

\begin{equation}
Z_f(\beta,R)=\frac{2\pi}{\beta\Gamma(f/2)}\gamma(\left(\frac{f}{2},
\frac{\beta R^2}{2}\right) ,
\end{equation}
with $\gamma(a,x)$ being the incomplete gamma function\cite{Abram}. 
The entropy of this distribution is 

\begin{equation}
S_q=\frac{1}{q-1}\left[1-\frac{Z_f(q\beta,R)}{Z_f^q(\beta,R)}\right] .
\end{equation}
Replacing the incomplete gamma function by its asymptotic 
expression\cite{Abram}
 
\begin{equation}
\gamma(a,x)=\frac{\exp(-x)x^{a}}{a} 
\end{equation}
the entropy assumes the correspondent asymptotic expression
 
\begin{equation}
S_q=\frac{1}{q-1}\left[1-\left(\pi R^2\right)^{(1-q)f/2}
\Gamma^{q-1}(1+f/2)\right]  .
\end{equation}
If in this expression, the radius $R$ is made dependent on the degree of freedom 
$f$ as
 
\begin{equation}
\pi R^2=(\frac{f}{2}+1) \exp(-1)  \label{57j}
\end{equation}
and the gamma function is replaced by its Stirling approximation
 
\begin{equation}
\Gamma(z)=\sqrt{2\pi}z^{z-\frac{1}{2}}\exp(-z),
\end{equation}
the $S_q$ becomes 
 
\begin{equation}
S_q=\frac{1}{q-1}\left(1-\left[2\pi(\frac{f}{2}+1)\right]^
{\frac{q-1}{2}}\exp(1-q)\right)  .
\end{equation}
Therefore for the value  $q=3$ of the entropic parameter, $S_3$
increases linearly with $f$ as

\begin{equation}
S_3=\frac{\pi}{2e^2}f . \label{57h}
\end{equation}
This is illustrated in the figure where $S_q$ is plotted as function 
of $f$ for the values 2,3 and 4 of the entropic parameter.  

To investigate how the restriction in the occupation of the phase
space affects the scale-invariant property of the joint 
density probability distribution, let us integrate it over one
variable of the set of $f$ variables. As the variables appear
equivalently in the distribution, we take by convenience the last 
one, i.e. $h_f,$ integrating it over the available domain, 
we obtain the distribution 

\begin{equation}
P_{f}^{\prime}(h_1,...,h_{f-1})=Z_f^{-1}\exp\left(-\frac{\beta}{2}
\sum^{f-1} _{i=1} h^2_i\right) 
2\int_{0}^{\sqrt{R^2-\sum^{f-1} _{i=1}h^2_i}}dh_f\exp(-\frac{\beta}{2}h_f^2) .
\end{equation}
Using the asymptotics of the functions we get

\begin{equation}
P_{f}^{\prime}(h_1,...,h_{f-1})=\left(\frac{\beta}{2\pi}\right)^{f-1/2}
\frac{\exp\left[-\frac{\beta}{2}\left( R^2+\sum^{f-1} _{i=1} h^2_i
\right)\right]
\mbox{erf}\left(\beta\sqrt{R^2-\sum^{f-1} _{i=1}h^2_i}/2\right)}
{\left(\pi R^2\right)^{-f/2}\Gamma(1+f/2)} .
\end{equation}
At the bulk of the space when $f\rightarrow\infty$ the error function
tends to one. Taking the ratio of this distribution with the
distribution of for $f-1$ after replacing the gamma functions by its
Stirling approximation we obtain

\begin{equation}
\lim_{f\rightarrow\infty}
\frac{P_{f}^{\prime}(h_1,h_2,...,h_{f-1})}{P_{f-1}(h_1,h_2,...,h_{f-1})}
=\sqrt{\frac{\beta}{2\pi}} .
\end{equation}

We remark that from Eq. (\ref{57j}), $R\sim \sqrt{f},$ such that 
the restricted domain occupied by the distribution expands
concomitantly with the expansion of the phase
space. Therefore, asymptotically,  in the
limit $f\rightarrow\infty, $ the whole phase space is occupied.

\subsection{Discrete case}

Consider now the joint distribution of matrix elements of the 
adjacent matrix of the ER model. Following the same scheme we impose 
the condition that the the trace can not be greater than a value $d,$ 
that is the variations of the set $a$ of random binary variables 
are constrained to obey the condition

\begin{equation}     
\sum_i^f a_i=k \leq d.
\end{equation}
The probability that the sum be equal to $k$ has to be redefined
as

\begin{equation}
P_f (\alpha,k)=\left\{
\begin{array}{rl}
\frac{\exp(-\alpha k)}{[1+\exp(-\alpha)]^f}, &\mbox{if } 
f\leq d \\
\frac{\exp(-\alpha k)}{W_f[1+\exp(-\alpha)]^f}, &\mbox{if } 
 f>d \mbox{ and }k\leq d \\
0, &\mbox{if } f>d \mbox{ and } k>d
\end{array}
\right.                     .   \label{17d}
\end{equation}
In (\ref{17d}), the normalization constant is given by

\begin{equation}
W_f (\alpha,d)=\sum_{k=0}^d \frac{f!}{k!(f-k)!}
\frac{\exp\left(-k\alpha \right)}{[1+\exp(-\alpha)]^f}  . \label{19a}
\end{equation}
The entropy of this distribution is

\begin{equation}
S_q=\frac{1}{q-1}\left(1-\frac{1}{W_{f}^{q}}\sum_{k=1}^d 
\frac{f!}{k!(f-k)!} \frac{\exp\left(-qk\alpha \right)}
{[1+\exp(-\alpha)]^{qf}}\right) . \label{19b}
\end{equation}
For $f>>d,$ the sums in (\ref{19a}) and (\ref{19b}) can be
approximated by their last terms as 

\begin{equation}
W_f (\alpha,d)\sim \frac{f^d}{d!}
\frac{\exp\left(-d\alpha \right)}{[1+\exp(-\alpha)]^f}  
\end{equation}
and 

\begin{equation}
\sum_{k=1}^d 
\frac{f!}{k!(f-k)!} \frac{\exp\left(-qk\alpha \right)}
{[1+\exp(-\alpha)]^{qf}} \sim \frac{f^d}{d!}
\frac{\exp\left(-qd\alpha \right)}{[1+\exp(-\alpha)]^{qf}},  
\end{equation}
which substituted in  (\ref{19b}) lead to the approximated expression 

\begin{equation}
S_q\sim \frac{1}{q-1}\left(1-\frac{f^{(1-q)d}}{(d!)^{1-q}}\right) 
\end{equation}
for the entropy. Immediately we conclude that making

\begin{equation}
q=1-\frac{1}{d} 
\end{equation}
$S_q$ becomes extensive in the limit $f\rightarrow\infty.$

To discuss the scale-invariant property of the distribution  
(\ref{17d}), we first note that apart from the normalization
$W_f (\alpha,d)$ the expressions are the original probabilities 
of the Erd\"{o}s-Renyi graph which satisfy the scale-invariant property.
Therefore, it is enough to consider that extra term, that is the
ratio between its value when $f$ changes to $f+1.$ Using its
approximation by the last term we obtain

\begin{equation}
\frac{W_{f+1} (\alpha,d)}{W_{f} (\alpha,d)}\sim 
\frac{1}{1+\exp(-\alpha)}\left(\frac{f+1}{f}\right)^d ,
\end{equation} 
which using the relation (\ref{57}) becomes

\begin{equation}
\frac{W_{f+1} (\alpha,d)}{W_{f} (\alpha,d)}\sim 
(1+p)\left(\frac{f+1}{f}\right)^d .
\end{equation} 
Assuming that the probability scales as $p\sim N^{-z},$ we conclude
that the distribution satisfies asymptotically the scale-invariant
property. 

\section{Conclusion} 

We discuss the extensivity of the nonadditive entropy, $S_q,$
using the joint density distribution of certain random matrix ensembles. This
entensivity has  previously been investigated, using artificial
distributions constructed with that objective\cite{Sato}. Here we are
studying it resorting to probability distributions of statistical
models with applications in many area. 
Our main result is to set the role the restriction in
the variations of the random variables plays, in order to have 
extensivity for values different from one of the entropic
parameter. We also verified that when this happens asymptotically 
the scale-invariant property condition is satisfied. Of course,
restriction in the occupation of phase space 
concomitantly entails correlations among the variables.

This work is supported by the Brazilian agencies CNPq and FAPESP.

{\bf Figure Captions}

Fig 1. Eqs. (\ref{35s}) and (\ref{35t}) for the exact and the
asymptotic expressions of the $S_q$ entropy of the disordered 
($\beta=1$) Gaussian ensemble are plotted, as a
function of the number of variables $f,$ fixing 
$\frac{1}{q-1}+\frac{f}{2}={\bar \xi}$  with ${\bar \xi}=4.$  

Fig 2. The expression of the $S_q$ entropy of the disordered 
adjacency matrices is plotted, as a
function of number of variables $f,$ fixing $\frac{1}{q-1}+\frac{f}{2}=\xi$  
with $\xi=6.$ 

Fig 3. The $S_q$ entropy of the restricted ($\beta=1$) Gaussian ensemble 
is plotted as a function of  the number of variables $f$ for 
$q=2,3$ and $4,$ The dashed line is the approximating asymptotic 
expression Eq. (\ref{57h}). 

\end{document}